\documentclass[sn-mathphys, icol]{sn-jnl}
\usepackage{subcaption}
\usepackage{siunitx}
\DeclareSIUnit\gauss{G}

\begin{document}
	\title{Dynamic self-organisation and pattern formation by magnon-polarons}
	
	\author[1,2]{M.~Gidding}
	\author[1,2]{T.~Janssen}
	\author[1,2]{C.~S.~Davies}
	\author[1,2]{A.~Kirilyuk}
	\affil[1]{FELIX Laboratory, Radboud University, Toernooiveld 7, 6525 ED Nijmegen, The Netherlands}
	\affil[2]{Radboud University, Institute of Molecules and Materials, Heyendaalseweg 135, 6525 AJ Nijmegen, The Netherlands}

	\abstract{Nonlinear dynamics can give rise, via the processes of self-organisation and pattern formation, to the spontaneous manifestation of order in open and complex systems far from equilibrium.
	Self-organising systems, transforming the inflow of energy into information,
	are ubiquitously found in current topical areas of science ranging from brainwave entrainment and neuromorphic computing to energy-efficient data storage technologies.
	In the latter, magnetic materials play a pivotal role combining very fast switching with permanent retention of information. However, it has been shown that, at very short time scales, magnetisation dynamics become chaotic due to internal instabilities, resulting in incoherent spin-wave excitations
	that ultimately destroy magnetic ordering. Here, contrary to all expectations, we show that such chaos gives rise to a periodic pattern of reversed magnetic domains, with a feature size far smaller than the spatial extent of the excitation. We explain this pattern as a result of phase-synchronisation of magnon-polaron waves, driven by strong coupling of magnetic and elastic modes. Our results reveal not only the peculiar formation and evolution of magnon-polarons at short time-scales, but also present a novel mechanism of magnetization reversal driven by coherent packets of short-wavelength quasiparticles.}

	\maketitle
	
	Finding methods that enable fast, efficient and long-lasting reversal of magnetisation direction represents a major topic of research in condensed matter physics with obvious technological applications~\cite{10.1007/978-3-319-09228-7_4}. The most theoretically-straightforward but also practical approach for switching magnetisation involves very-large-amplitude precessional motion~\cite{Gerrits2002,PhysRevLett.90.017201,Tudosa2004}. 
	However, the dynamics of such magnetisation reorientation across a large angle are dramatically different from the magnetisation precession in the regime of ferromagnetic resonance (FMR)~\cite{PhysRevLett.90.167203}. Very fast reorientations of magnetisation are inexorably accompanied by instabilities~\cite{Tudosa2004,PhysRevLett.96.047601} deriving from the spin-wave instability of overexcited FMR (the ``Suhl instability''). It is well understood that the Landau-Lifshitz-Gilbert equation still characterises such motion but the effective dissipation constant considerably exceeds, by order(s) of magnitude, that which is typical of FMR~\cite{PhysRevLett.94.197603}. Such damping is related to the excitation of a large spectrum of spin waves during the precession~\cite{PhysRevLett.96.047601,SUHL1957209,https://doi.org/10.1002/pssb.2220920124,https://doi.org/10.1002/pssb.2220930106,PhysRevLett.90.167203}. The faster the magnetic switching is driven, the more energy is pumped into the system, and therefore the more of it flows into incoherent spin-wave modes with wave vectors in very broad range~\cite{Bauer2015}. 
        
    In addition to the non-linear instabilities arising from the magnetostatic interactions within the magnetic system, an extra channel of non-linearity is provided by the interaction of the excitation with the lattice~\cite{TURITSYNSK1985SOMS}. The non-linearity and dispersion of acoustic waves, for example, can be amplified by the presence of small-amplitude magnetisation oscillations at similar frequencies. It is completely unknown, however, how such coupling influences magnetisation switching featuring a large cone angle of precession.
        
    Here, we use multiscale pump-probe experiments, spanning timescales ranging from nanoseconds to milliseconds, to demonstrate how spin-wave instabilities result in distinct pattern formation, in the process of switching of magnetisation. Utilising infrared single pulses derived from a cavity-dumped free-electron laser\cite{janssen2022cavity}, we strongly pump optical phonons at resonance, dynamically deforming the structure of the material~\cite{Mankowsky_2016}. This temporally creates a magnetic anisotropy field that drives, in turn, large-amplitude dynamics and switching of the magnetisation. The large-amplitude precessional motion leads to strong instability of the spin system and generates a sizeable number of spin waves, strongly populating the low-energy part of the magnon spectrum. The subsequent repopulation in the presence of a magneto-elastic interaction leads to effective `condensation' of the waves into the magnon-polaron part of the spectrum, observed both experimentally and in micromagnetic simulations~\cite{PhysRevLett.118.237201,shen2018theory}. Ultimately, these magnon-polaron waves synchronize and achieve such high amplitude that they induce complete switching of magnetisation at the wave maxima.  

		\begin{figure}
			\centering
			\begin{subfigure}{0.24\linewidth}
				\includegraphics[width=\linewidth]{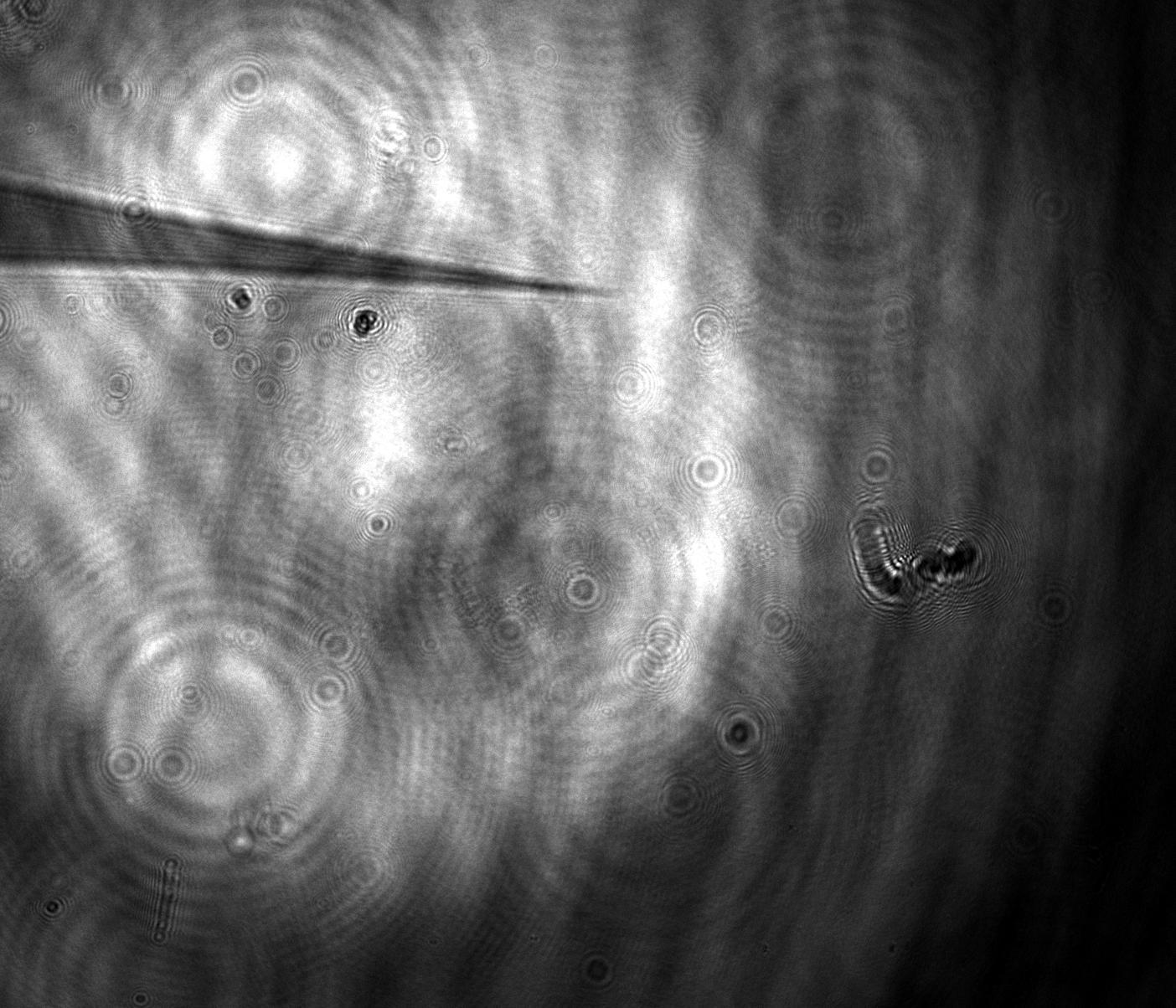}
				\caption{}
			\end{subfigure}
			\begin{subfigure}{0.24\linewidth}
				\includegraphics[width=\linewidth]{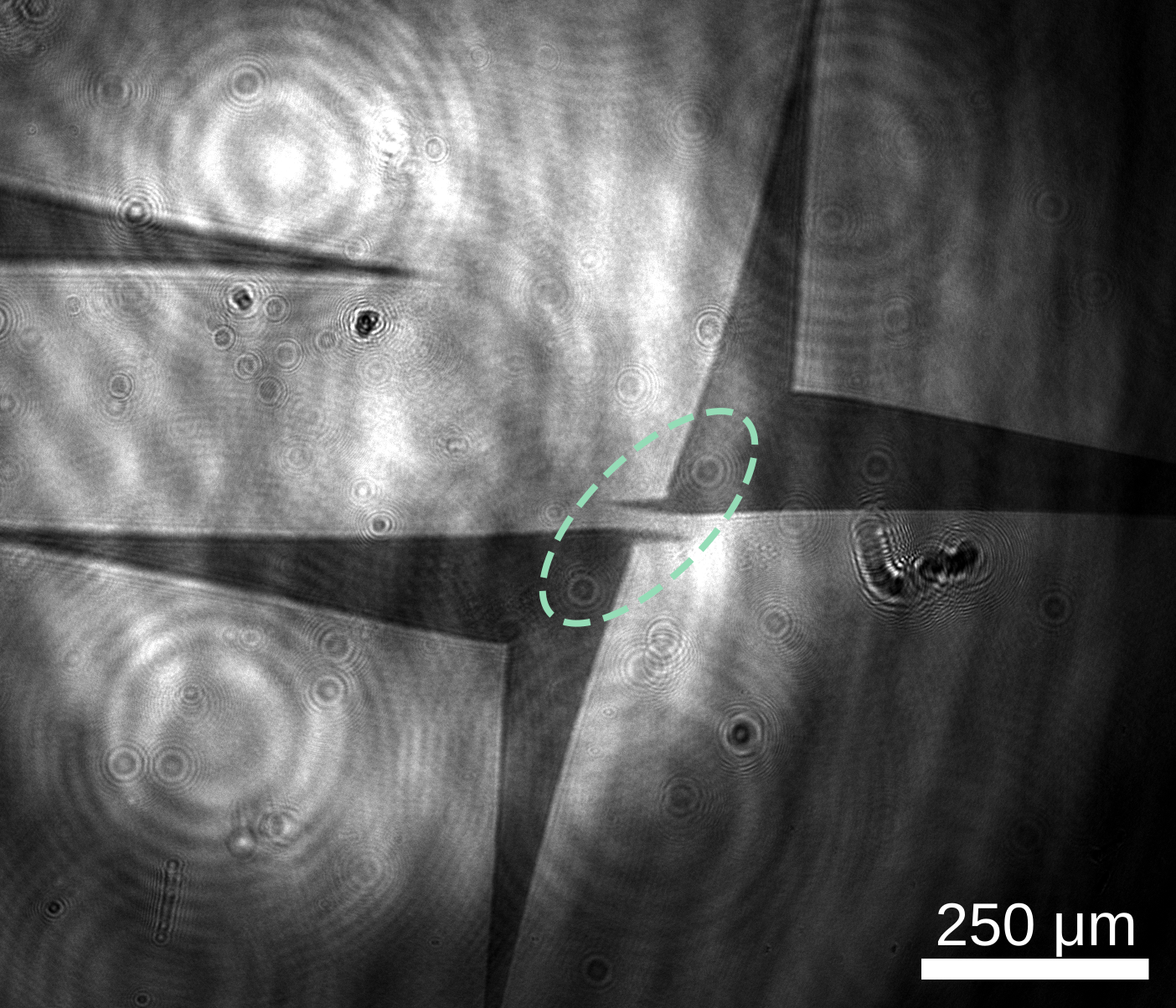}
				\caption{}
			\end{subfigure}
			\begin{subfigure}{0.24\linewidth}
				\includegraphics[width=\linewidth]{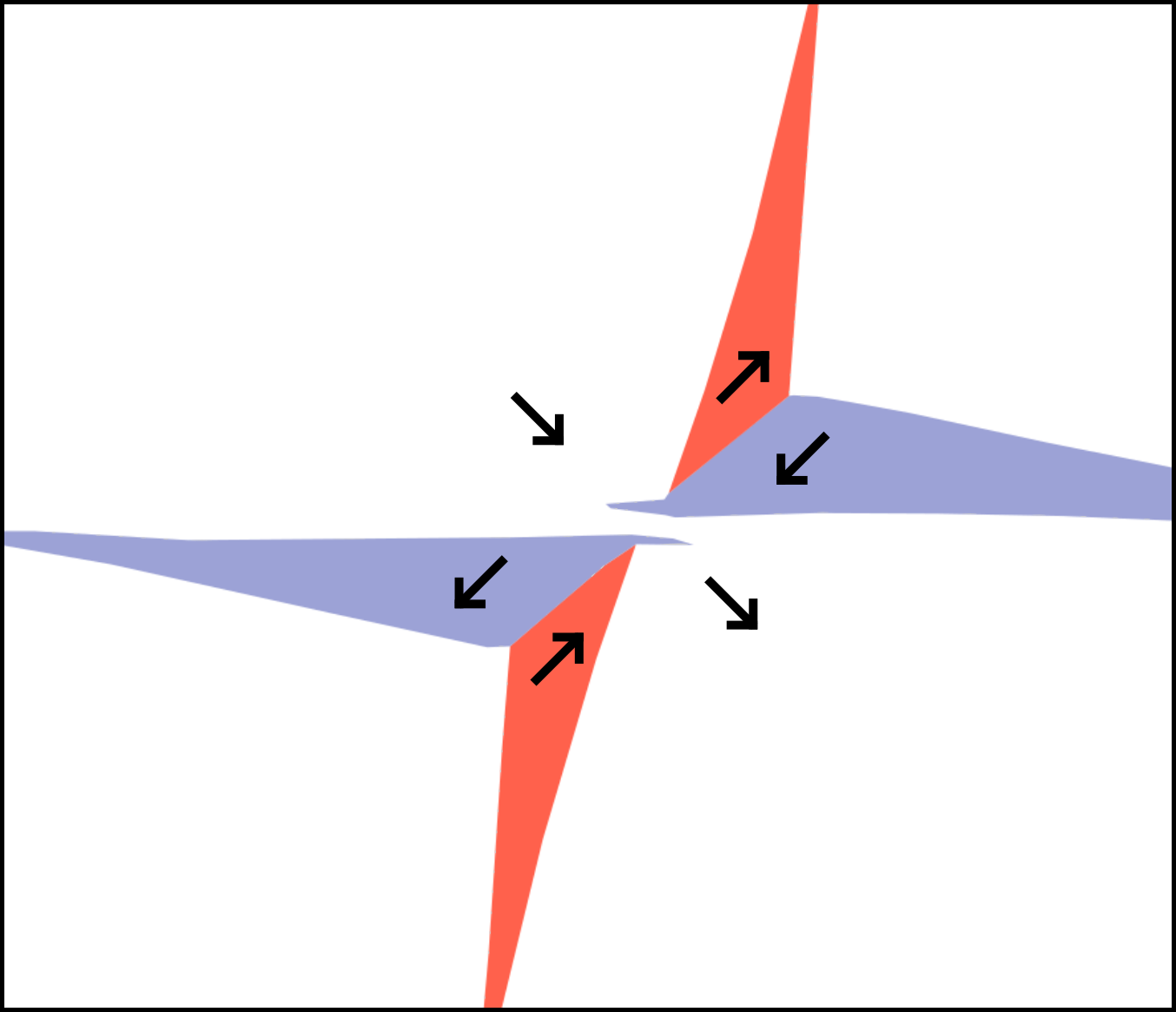}
				\caption{}
			\end{subfigure}
			\begin{subfigure}{0.24\linewidth}
				\includegraphics[width=\linewidth]{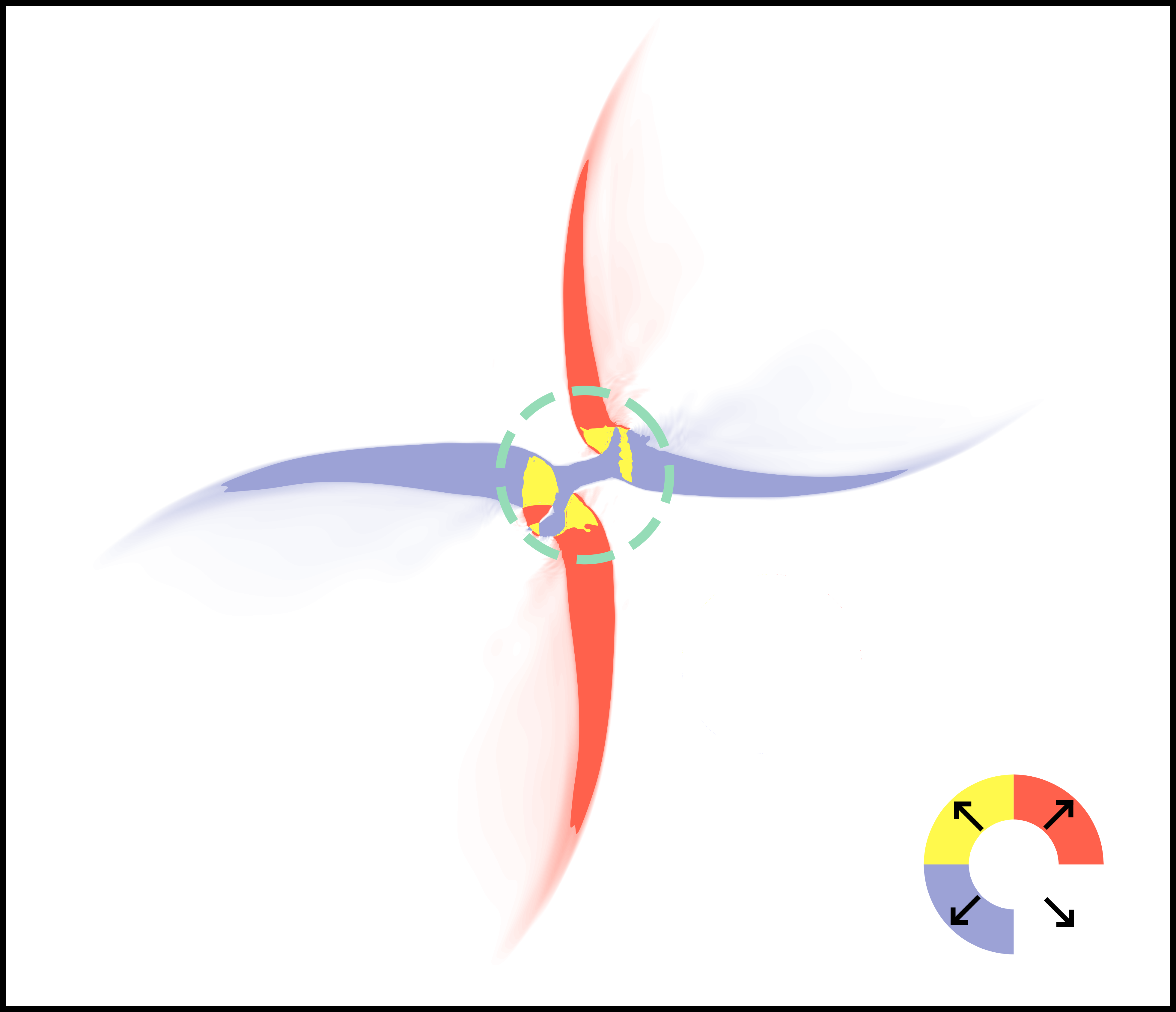}
				\caption{}
			\end{subfigure}
			\caption{\textbf{Phononic switching of in-plane magnetisation in Lu:YIG.} Images of the magnetisation distribution taken \textbf{a} before and \textbf{b} \SI{200}{\micro \second} after irradiation by a single pump pulse of wavelength \SI{13}{\micro \meter} and duration of \SI{\approx 2}{\pico \second}. \textbf{c} The switched magnetic domains isolated by subtracting the non-magnetic background. \textbf{d} The direction of the magnetisation vector in the plane of the simulated sample, after a pulse strikes the sample. In panels \textbf{b} and \textbf{d}, the dashed line indicates the half-maximum of the pump pulse.}
			\label{switching}
		\end{figure}

	In our study, we used a magnetic \SI{7.5}{\micro\meter}-thick lutetium- and bismuth-doped yttrium-iron-garnet film, hereafter referred to as Lu:YIG~\cite{PhysRevLett.95.047402}. This sample has a weak in-plane fourfold anisotropy field of $\SI{4}{\kilo \ampere \per \meter}$ which results, in the absence of magnetic fields, in four equilibrium orientations of the magnetisation that are orthogonal to each other in the plane of the film. We perform our experiments at the free-electron laser facility FELIX in the Netherlands~\cite{felix1995}. In the pump-probe measurements, we pump the sample with single transform-limited micropulses~\cite{janssen2022cavity} with a wavelength of \SI{13}{\micro \meter}, an energy of \SI{\approx 80}{\micro \joule}, and typical duration of $\approx$\SI{2}{\pico\second}. These pulses were focused on to the surface of the Lu:YIG sample to an elliptical spot with full-width-half-maximum diameter of 300 and \SI{130}{\micro \meter}~\cite{Liu:82}. The spatial distribution of magnetisation dynamics was magneto-optically probed using defocused pulses of wavelength \SI{532}{\nano\meter} and duration \SI{5}{\nano\second}, with the latter defining the temporal resolution of the experiment. To increase the observed contrast between the in-plane magnetisation directions of Lu:YIG, the sample was tilted by $\approx$30$^{\circ}$ relative to the probe's path.
	
    After exposing the sample to a single pump pulse, we observe that the homogeneous magnetisation distribution switches to form a peculiar spatial pattern (Fig.~\ref{switching}a-b). The switching consists of four distinct, triangular domains that emanate diagonally outwards from the center of the irradiated region. At the center, however, switching is absent. By subtracting the images taken before and after illumination, we clearly identify that the domains have two-fold rotational symmetry (Fig.~\ref{switching}c). 
    
    The observed four-domain pattern results from large-amplitude magnetisation dynamics~\cite{Stupakiewicz2021} and can be understood in terms of the magnetoelastic interaction. Specifically, the resonant pumping of infrared-active optical phonon modes microscopically displaces the equilibrium atomic positions along an optical phonon coordinate~\cite{PhysRevB.89.220301}. The effect of such deformation on magnetisation can be modelled using a standard micromagnetic description of magnetoelastic energy~\cite{Stupakiewicz2021}. Indeed, simulations of Lu:YIG with the inclusion of the magnetoelastic interaction successfully reproduce the switching pattern observed experimentally, as shown in Fig.~\ref{switching}d. The spectral dependence of the switching (see Supplementary Materials) confirms the phononic mechanism.
    	
\begin{figure}
		\centering
		\begin{tabular}{ccc}
    	 	\SI{-75}{\nano \second} & \SI{1}{\nano \second} & \SI{5}{\nano \second} \\
    	    \includegraphics[width=0.30\linewidth]{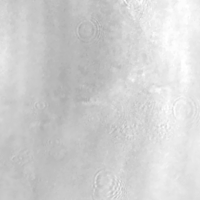} &
    		\includegraphics[width=0.30\linewidth]{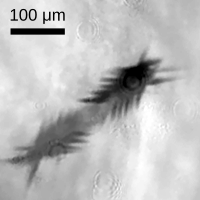} & \includegraphics[width=0.30\linewidth]{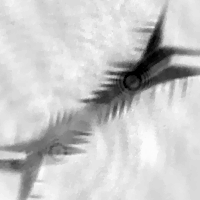} \\
    	 	\SI{15}{\nano \second} & \SI{55}{\nano \second} & \SI{105}{\nano \second} \\
    	 	\includegraphics[width=0.30\linewidth]{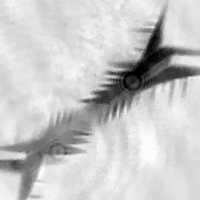} & \includegraphics[width=0.30\linewidth]{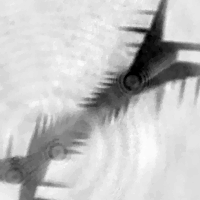} & \includegraphics[width=0.30\linewidth]{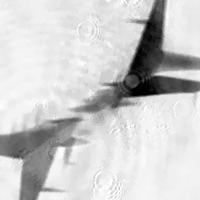} \\
    	 	\SI{155}{\nano \second} & \SI{305}{\nano \second} & \SI{1005}{\nano \second} \\
     	 	\includegraphics[width=0.30\linewidth]{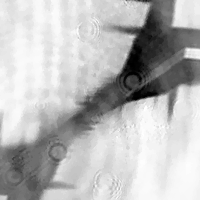} &
    	 	\includegraphics[width=0.30\linewidth]{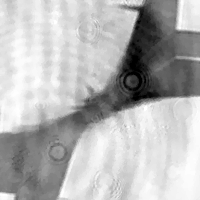} &  \includegraphics[width=0.30\linewidth]{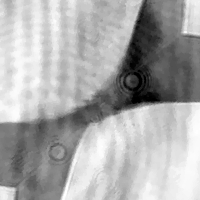}\\
		\end{tabular}
		\caption{\textbf{Time-resolved imaging of the pattern formation.} \textbf{a}-\textbf{i} Representative background-subtracted images taken at the indicated times after the arrival of the pump pulse (wavelength \SI{13}{\micro \meter}, duration $\approx$\SI{2}{\pico \second}). 
		}
		\label{ns}
	\end{figure}

    We now proceed to single-shot pump-probe measurements, varying the time of arrival of the visible probe with respect to the pump, to understand how the pattern of switching forms. Typical results of this experiment are shown in Fig.~\ref{ns}. These results show that the triangular domains shown in Fig.~\ref{switching} propagate outwards from the center of irradiated spot, continually growing in size. More intriguingly, however, during the nascent stages of the pattern formation, we observe a periodic pattern of domains (hereafter referred to as ``ripples'') with a visible period on the order of 10-\SI{20}{\micro \meter}, which is very much unexpected given the large ($\sim \SI{300}{\micro \meter}$) diameter of the excitation spot along this axis. During the first hundreds of nanoseconds following the arrival of the pump pulse, Fig.~\ref{ns} shows that the pattern is remarkably heterogeneous with ripples propagating outwards from the center. These magnetisation ripples are very reproducible and appear for other orientations of magnetisation in the sample as well (see the Supplementary Materials). While the large triangular domains persist for a timescale of hundreds of microseconds to milliseconds (see Supplementary Materials), the ripples have decayed completely within about \SI{300}{\nano\second} after the arrival of the pump pulse.

    \begin{figure}
        \centering
		\includegraphics[width=0.7\linewidth]{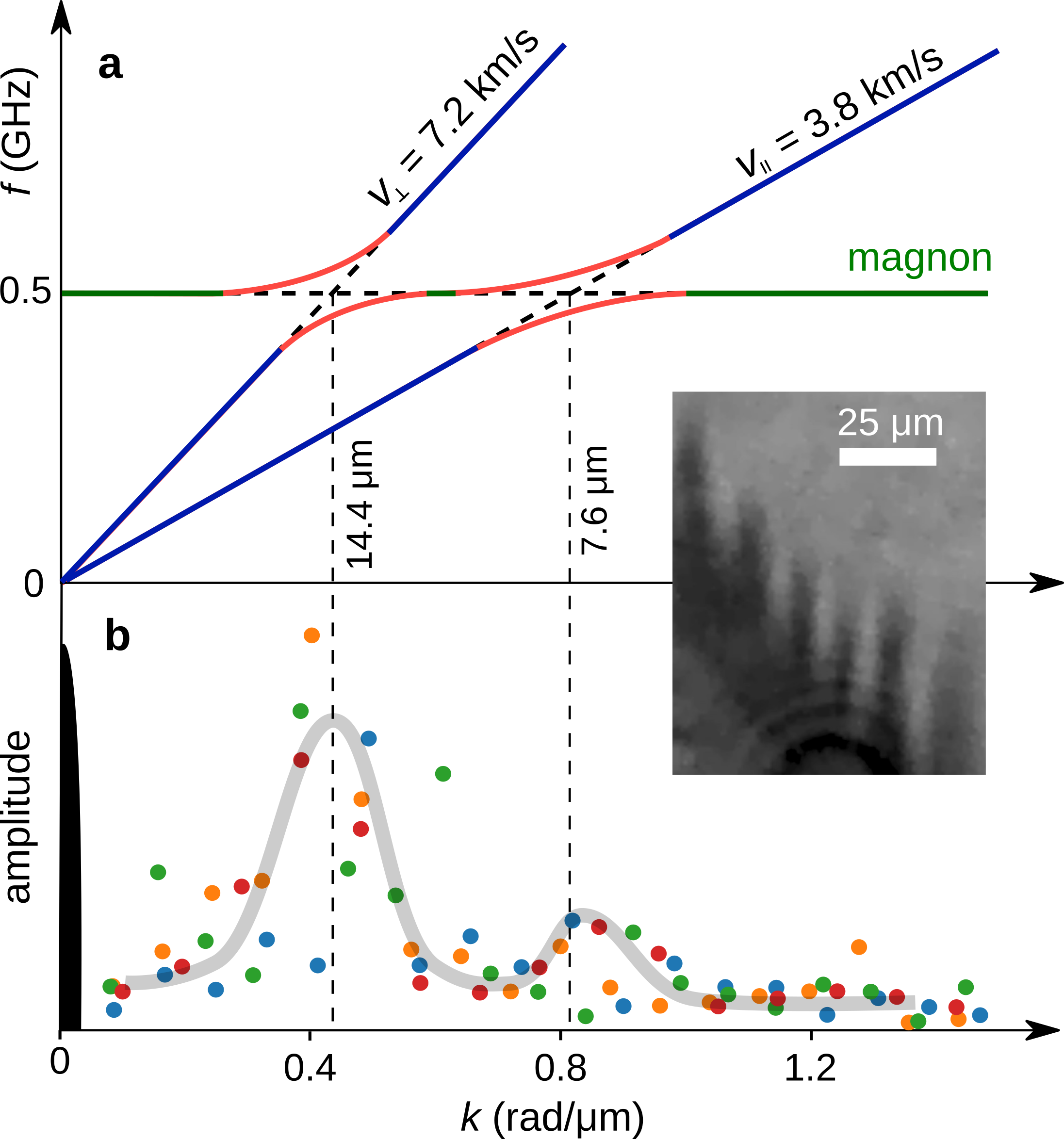}
		\caption{\textbf{Schematic illustration of the self-accumulation.} \textbf{a} Sketch of the anti-crossings (red) between the acoustic phonons (blue) and the magnon (green). \textbf{b} Amplitude of the Fourier transform of the ripple patterns during the first \SI{40}{\nano \second} (coloured circles) and of the pump pulse (black), with a guide to the eye (grey). Inset: zoomed section of the magneto-optical image of the ripple pattern used for the Fourier transform.}
		\label{dispersion}
	\end{figure}
    
    It is well understood that a spatially-localised perturbation creates propagating waves with wave vectors determined by the profile of the excitation~\cite{PhysRevLett.115.197201,shen2018theory}. The results shown in Fig.~\ref{ns} therefore raise an intriguing question of how an excitation that is reasonably homogeneous (across $>$\SI{200}{\micro \meter}) results in such a heterogeneous ripple pattern with periodicity on the order of 10-\SI{20}{\micro \meter}. One possible explanation could be provided by the Faraday waves that appear predominantly in fluids subject to a periodic vertical oscillation\cite{doi:10.1098/rstl.1831.0018}, but which are also possible in solids~\cite{bevilacqua2020faraday}. However, taking into account the elastic parameters of YIG crystals together with the excitation frequency, we are able to rule out this possibility.
    
    Instead, we posit that such small features originate from spin-wave instabilities~\cite{SUHL1957209, https://doi.org/10.1002/pssb.2220920124, https://doi.org/10.1002/pssb.2220930106} in the presence of magnetoelastic interactions. The magnetisation precession is driven to such a high amplitude that the precessional motion of the magnetisation becomes highly unstable, leading to the generation of a large number of spin waves in a broad frequency spectrum~\cite{SUHL1957209,PhysRevLett.94.197603}. Such an excitation should therefore lead to chaotic magnetisation dynamics as shown in Ref.~\cite{PhysRevLett.96.047601}. However, instead we observe a rather well-defined quasi-periodic pattern of reversed domains, with a periodicity an order of magnitude smaller than what could be expected from the pulse profile. What can be the mechanism creating such a pattern?  

    Yttrium-iron-garnet films are renowned in the research field of magnonics for their low damping both in the elastic and magnetic subsystems~\cite{cherepanov1993saga,serga2010yig}. This combination of material properties yields access, via the strong coupling between the two subsystems, to regions in phase space where the magnon and phonon dispersion curves approach and consequently where their frequencies coincide. In these regions, the magnetoelastic coupling combines the magnons and phonons to form magnon-polarons~\cite{PhysRevLett.115.197201}. This quasiparticle hybridisation allows for the efficient transfer of angular momentum, as the slower magnons are dressed with phonons which have a significantly larger group velocity~\cite{PhysRevLett.115.197201, PhysRevB.99.060402}. This interaction between the magnetic and elastic subsystems, schematically illustrated in Fig.~\ref{dispersion}, can be observed directly by, for example, magneto-optics~\cite{PhysRevLett.105.117204,PhysRevLett.109.166601}.
    
    Brillouin scattering spectroscopy experiments have previously shown that the thermalisation of an overpopulated magnon gas, in combination with magnon-phonon scattering,  leads to the self-organisation of so-called magnetoelastic bosons at the crossing point of the magnon and phonon dispersions~\cite{PhysRevLett.118.237201}. This accumulation phenomenon is confirmed by micromagnetic simulations, taking into account the magnetoelastic interaction of magnon-polaron dynamics in Ref.~\cite{shen2018theory}. Calculations of spin pumping by a parametric excitation also prove that magnons are resonantly enhanced in strength at the magnetoelastic crossing point~\cite{PhysRevLett.121.237202}. Such a collapse of the broad magnon spectrum into the narrow range of magnon-elastic crossing can explain how, in our case, the broad spectrum resulting from spin-wave instabilities leads to the domain pattern with a well-defined periodicity.
    
    The above supposition is supported by a comparison of the periodicity of magnetic ripples observed during the first \SI{40}{\nano \second} after excitation (see Fig.~\ref{ns}) with what is expected from the $k$-vector of magnon-polarons. The ferromagnetic resonance frequency of our sample, given the very weak magnetic anisotropy and small saturation magnetisation, is about \SI{0.5}{\giga \hertz} in the case of magnetic fields close to zero \cite{Fredrik-PRB}. This, together with the sound velocities of \SI{3.8}{\kilo\meter\per\second} (\SI{7.2}{\kilo\meter\per\second}) for transverse (longitudinal) acoustic phonons~\cite{doi:10.1063/1.1736184}, leads to an estimated magnon-polaron wavelength of 8 to \SI{15}{\micro \meter}, which corresponds rather well to the smallest observed distances between the ripples. The fact that we see these waves as reversed domains supports the idea of `condensation' in the sense that the waves phase-synchronize with each other so that the large amplitude leads to the complete magnetisation reversal at the maxima. On the other hand, this phenomenon might even resemble neural entrainment, an observation that brain-waves (large-scale electrical oscillations in the brain) naturally synchronize to the rhythm of periodic visual, auditory or tactile stimuli~\cite{notbohm2016modification,thaut2015discovery,sieben2013oscillatory}.
    
    As an outlook, while our single-shot pump-probe imaging experiments clearly unveil the formation of large-amplitude magnon-polarons at the nanosecond time scale, further experiments must examine more closely the spatiotemporal details of their development at sub-nanosecond timescales. This could in particular provide a better insight into not only the intricacies of phonon-induced switching of magnetisation but also the dynamics of wave-synchronisation in general, which is a phenomenon of broad interest in different areas of science~\cite{thaut2015discovery,doveil2005experimental}.  
	
\bibliography{bibliography}

\section{Methods}
    \subsection{Materials}
    The sample studied is a \SI{7.5}{\micro\meter}-thick doped magnetic iron-garnet film with chemical composition Lu$_{1.69}$Y$_{0.65}$Bi$_{0.66}$Fe$_{3.85}$Ga$_{1.15}$O$_{12}$~\cite{PhysRevLett.95.047402}. The sample was grown on a (001) oriented gadolinium gallium garnet substrate with trace amounts of Pb impurities resulting from the growth process. This sample has a magnetisation of $M_s =\SI{300}{\kilo \ampere \per \meter}$~\cite{PhysRevB.70.094408} and a weak in-plane fourfold anisotropy field $\SI{4}{\kilo \ampere \per \meter}$ which results, in the absence of magnetic fields, in four equilibrium orientations of the magnetisation that are orthogonal to each other in the plane of the film. 
    
    \subsection{Methods}
    \subsubsection{Magneto-optical single-shot pump-probe microscopy experiments}
    To detect magnetisation dynamics of the samples, we use the technique of single-shot pump-probe microscopy, with the pump being delivered by FELICE at the free-electron laser facility FELIX in The Netherlands~\cite{felix1995}. Further details of this cavity-dumped laser are given in Ref.~\cite{janssen2022cavity}. This laser delivers transform-limited pulses with a central wavelength of \SI{13}{\micro \meter}, duration of $\approx$\SI{1}{\pico\second} and energy $\approx$\SI{80}{\micro \joule} at a rate of \SI{10}{\hertz}. A motorised shutter placed in the path of the infrared laser allows us to select single pulses. Using a 90$^{\circ}$ off-axis parabolic mirror, the infrared pump pulse is focused onto the surface of the Lu:YIG sample to an elliptical spot with a full-width-half-maximum diameter of 300 and \SI{130}{\micro \meter} as assessed by the Liu method~\cite{Liu:82}. The probe pulse is derived from a synchronised frequency-doubled Nd:YAG laser, which delivers linearly-polarised pulses of wavelength \SI{532}{\nano\meter} and duration $\approx$\SI{5}{\nano\second}. Upon transmission through the sample, the probe is gathered by the objective lens, passed through an analyser and detected using a camera. By detecting the rotation of polarisation of the light transmitted through the sample, induced by the magneto-optical Faraday effect, we are able to spatially resolve magnetisation dynamics across the sample. Electronically shifting the time of arrival of the probe pulse provides temporal resolution, limited to \SI{5}{\nano\second} as given by the probe's duration~\cite{janssen2022cavity}. To enhance the magneto-optical contrast between different states of the in-plane magnetisation of Lu:YIG, the sample was tilted at an angle of approximately 30$^{\circ}$ relative to the wave vector of the incoming probe pulse. 
    
    \subsubsection{Simulation}
    Micromagnetic simulations were performed using the MuMax$^3$ micromagnetic simulation program\cite{doi:10.1063/1.4899186}. The simulated sample dimensions were $10 \times 10 \times \SI{0.5}{\micro \meter}$ with cell size $4.88 \times 4.88 \times \SI{250}{\nano \meter}$, allowing for the inclusion of exchange interactions in the simulation. Literature values for physical constants in Lu:YIG were used, with the exchange stiffness set to $\SI{3.7}{\pico \joule \per \meter}$\cite{Klingler_2014} and the saturation magnetisation to $\SI{300}{\kilo \ampere \per \meter}$\cite{PhysRevB.70.094408}. A biaxial anisotropy was implemented in MuMax$^3$ via the method described by De Clercq \emph{et al.}\cite{De_Clercq_2017} with $K_b = \SI{750}{\joule \per \meter^3}$, and easy axes along the $(1, 1, 0)$, $(1, -1, 0)$ directions. In order to increase the demagnetising field strength to more closely match that of a thin film geometry, a uniaxial anisotropy with $K_u = \SI{-500}{\kilo \joule \per \meter^3}$ along the $(0, 0, 1)$ axis was added, creating an easy plane configuration in the sample plane which effectively mimics an in-plane demagnetising field geometry. Finally, the Landau-Lifshitz damping constant $\alpha$ was set to $0.02$, while at the outer 10\% of the simulated sample it increased quadratically up to $\alpha = 10$. This was done to create an absorbing boundary region, see Ref.~\cite{VENKAT201834} for a detailed description.
		
	Strain was introduced into the simulation using the method of Stupakiewicz \emph{et al.}\cite{Stupakiewicz2021}, which adds a transient modification to the strain tensor of the sample due to the non-linear interaction of phonon modes\cite{Mankowsky_2016}, mimicking the strain introduced by a Gaussian, infrared, laser pump pulse. The Gaussian standard deviation of the pulse was set to $\SI{0.333}{\micro \meter}$, equivalent to a full-width at half-maximum (FWHM) of $\SI{0.784}{\micro \meter}$, and the full-duration half-maximum (FDHM), $\tau$, to $\SI{10}{\pico \second}$. The peak of the Gaussian pulse strikes the sample $\SI{50}{\pico \second}$ after the simulation commences. 

\backmatter

\bigskip

\bmhead{Acknowledgments}

We acknowledge the Nederlandse Organisatie voor Wetenschappelijk Onderzoek (NWO-I) for their financial contribution, including the support of the FELIX Laboratory. We thank A.~V.~Kimel and A.~Stupakiewicz for discussions.

\clearpage

\begin{centering}

\Large{\textbf{Supplementary Material}} \\

\end{centering}

    \section{Time-resolved simulations}
\begin{figure}[h]
		\centering
		\begin{tabular}{ccc}
			\SI{-0.05}{\nano \second} & \SI{0.00}{\nano \second} & \SI{0.05}{\nano \second} \\
			\includegraphics[width=0.30\linewidth]{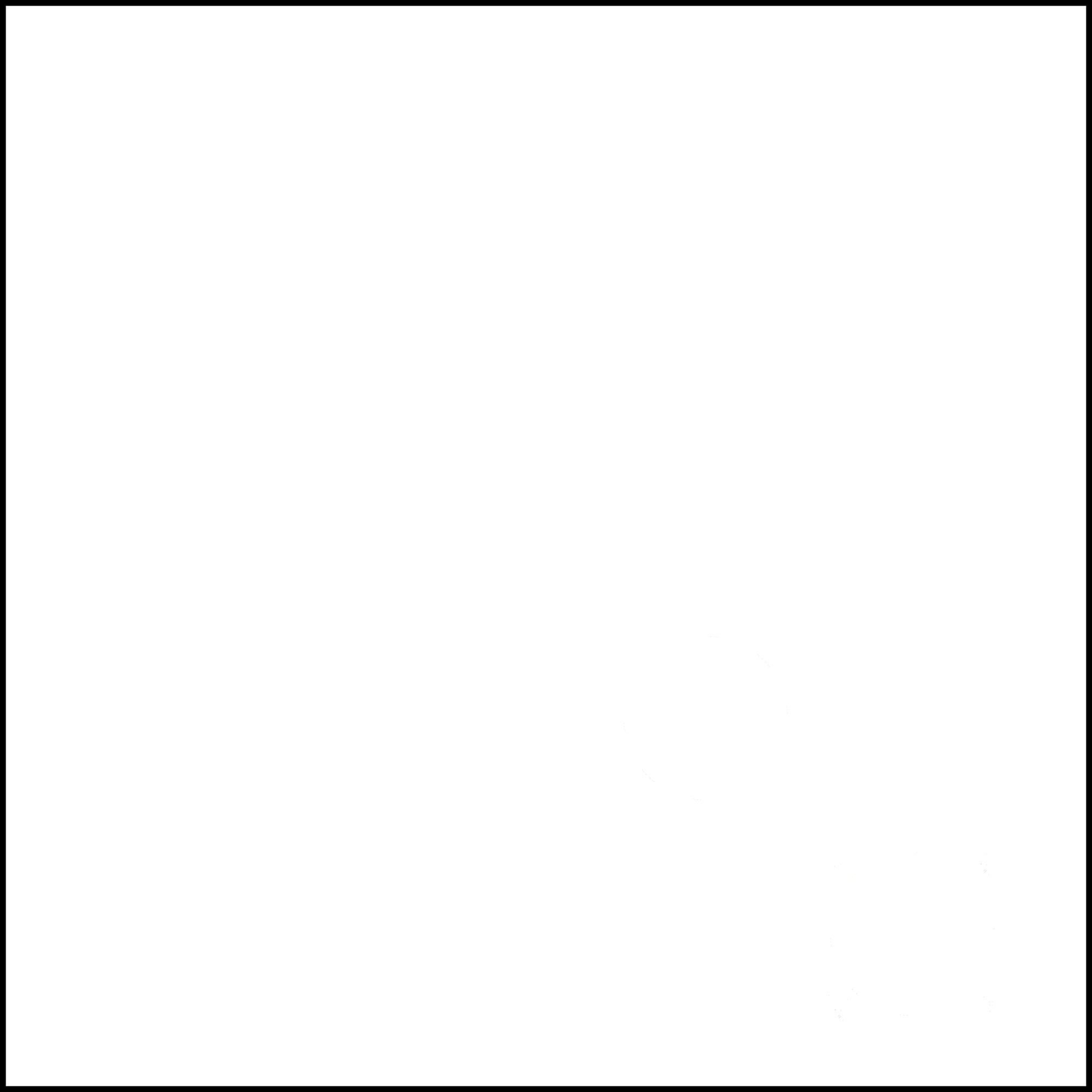} & \includegraphics[width=0.30\linewidth]{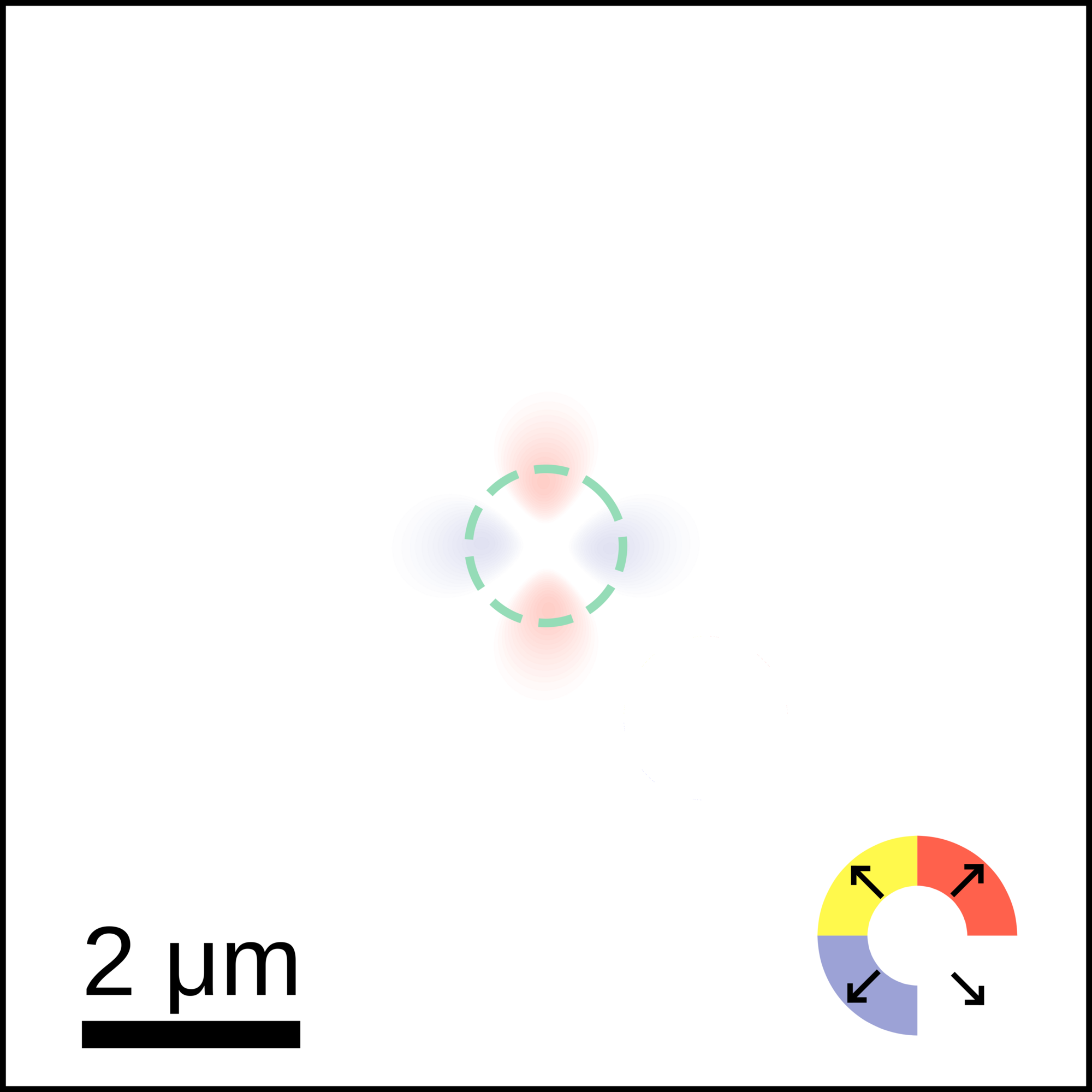} & \includegraphics[width=0.30\linewidth]{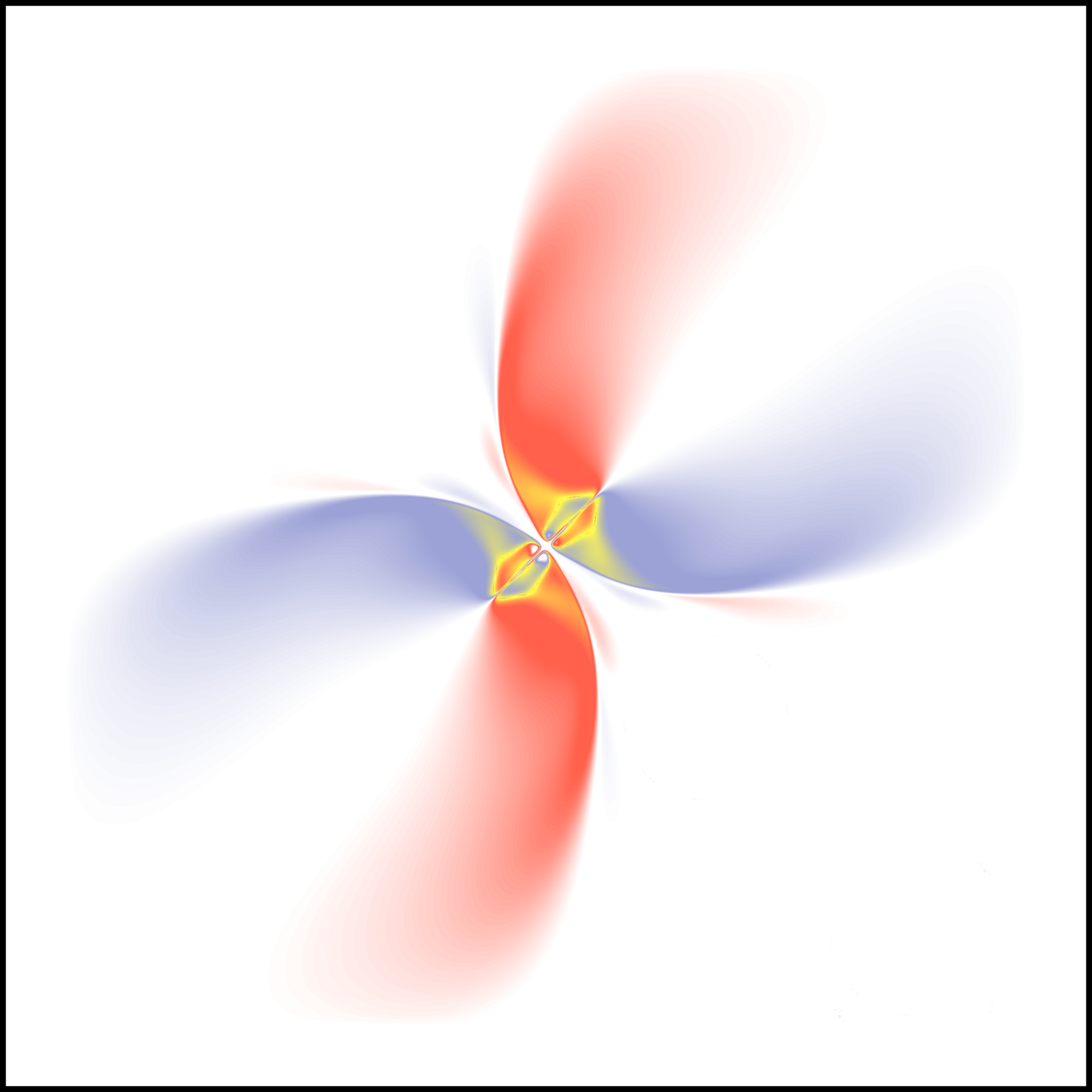}\\
		    \SI{0.10}{\nano \second} & \SI{0.15}{\nano \second} & \SI{0.30}{\nano \second} \\
		 	\includegraphics[width=0.30\linewidth]{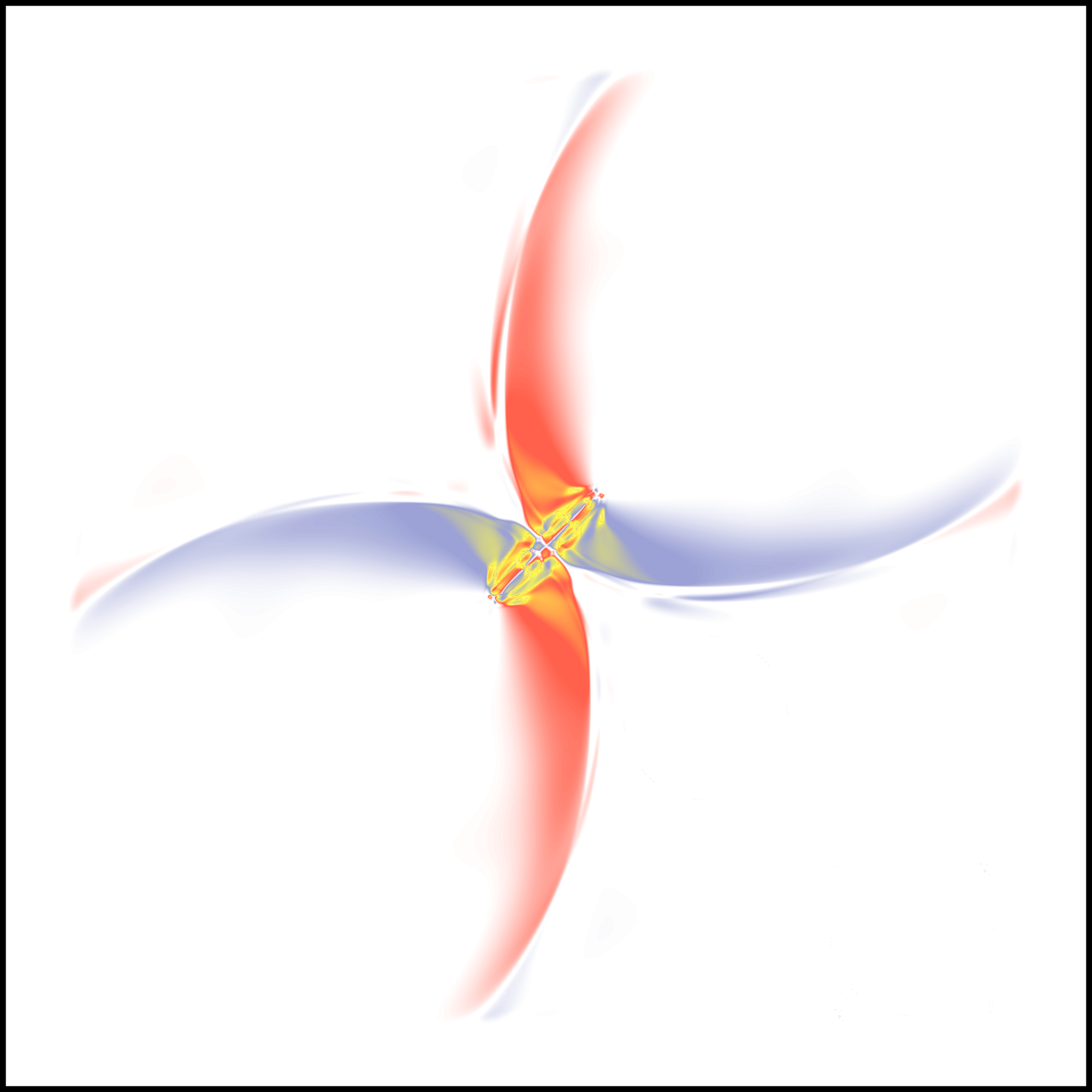} & \includegraphics[width=0.30\linewidth]{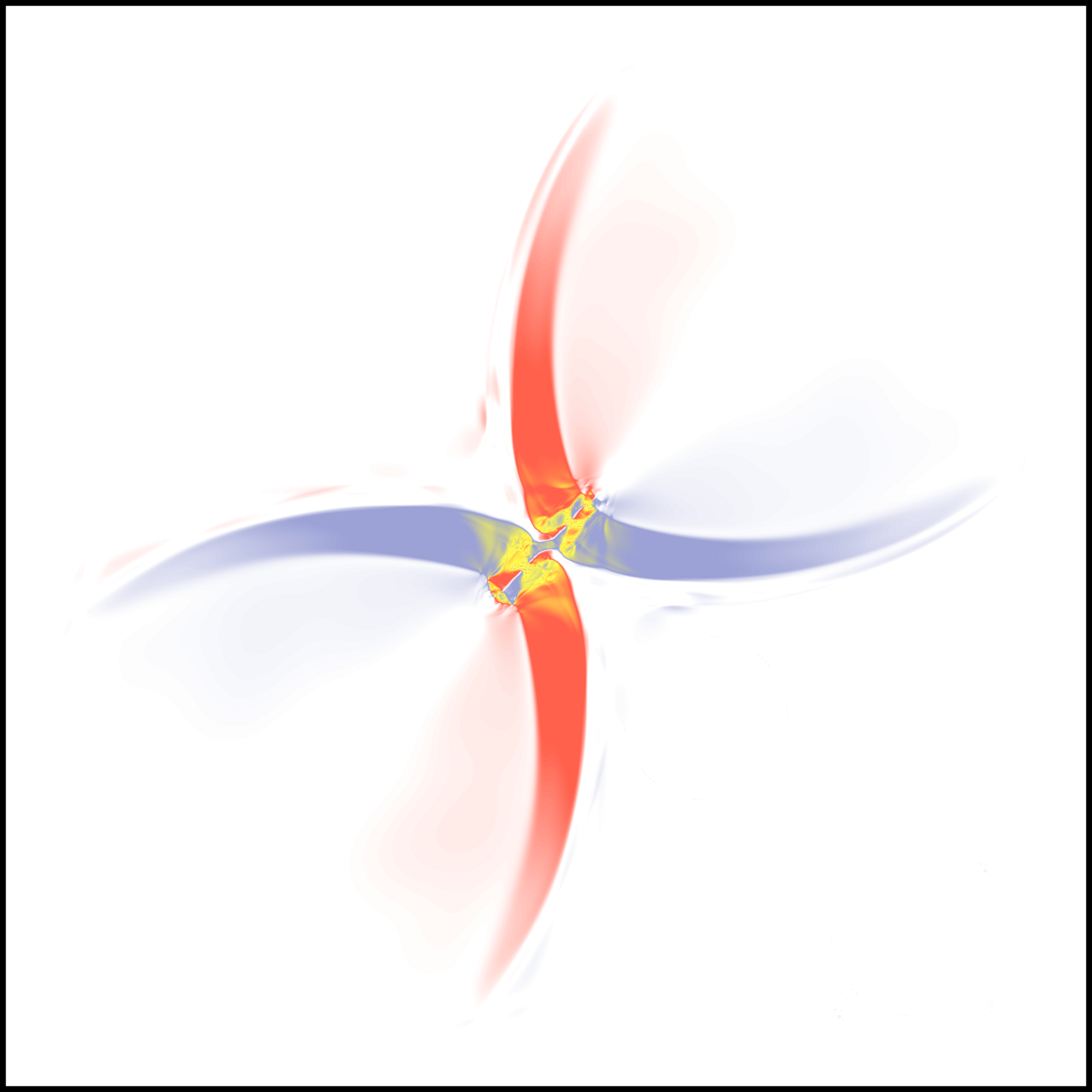} & \includegraphics[width=0.30\linewidth]{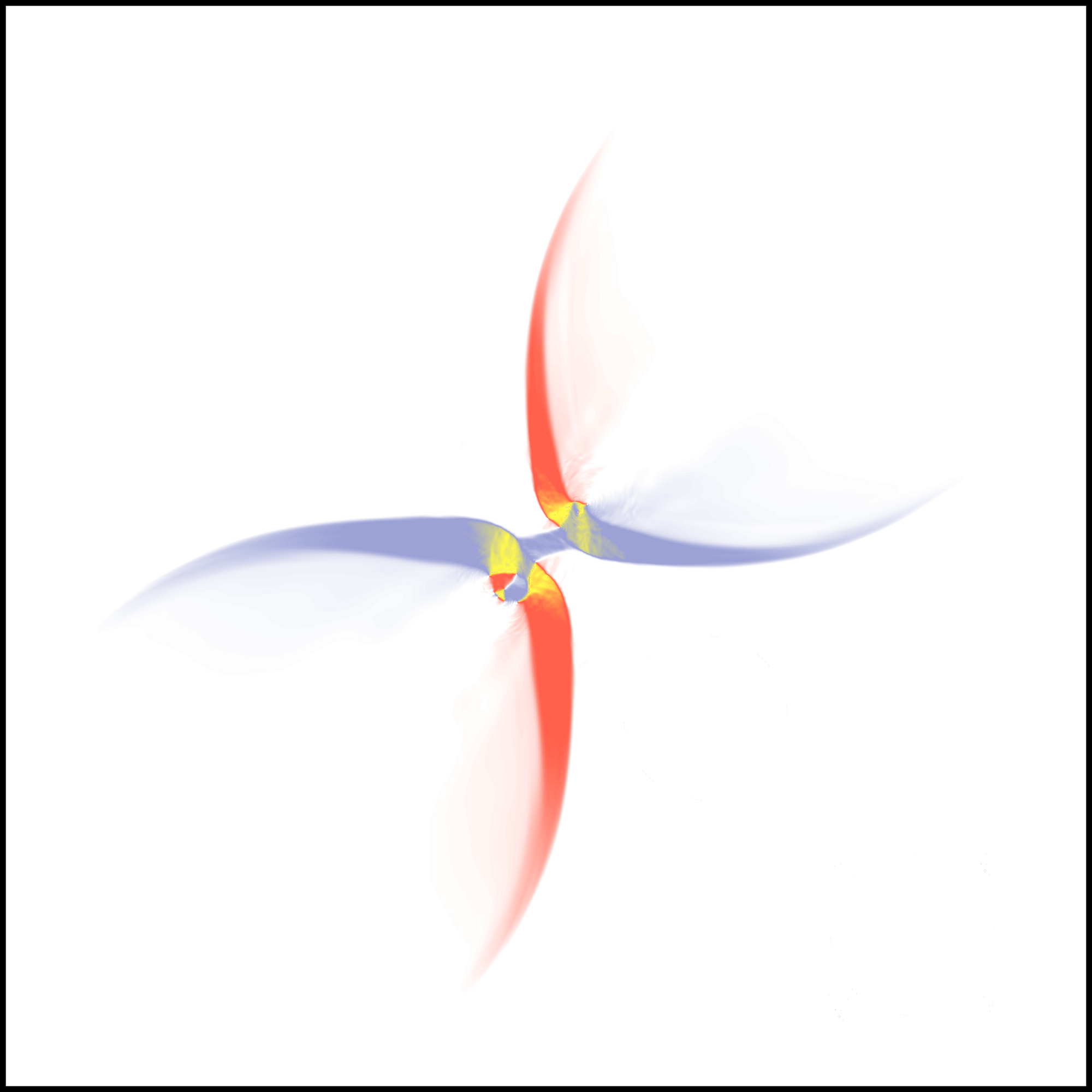}\\
		 	\SI{0.45}{\nano \second} & \SI{0.70}{\nano \second} & \SI{0.95}{\nano \second} \\
		 	\includegraphics[width=0.30\linewidth]{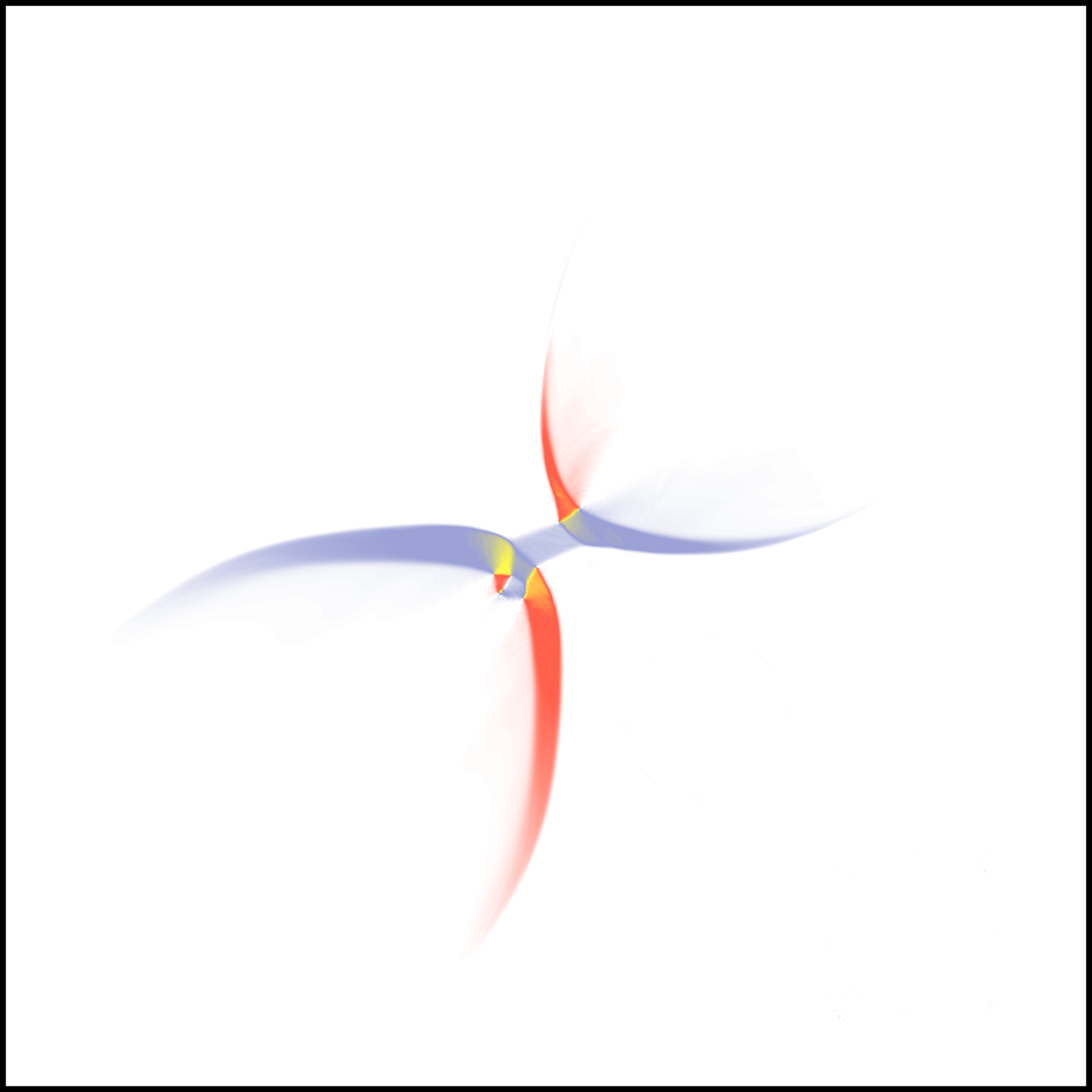} & \includegraphics[width=0.30\linewidth]{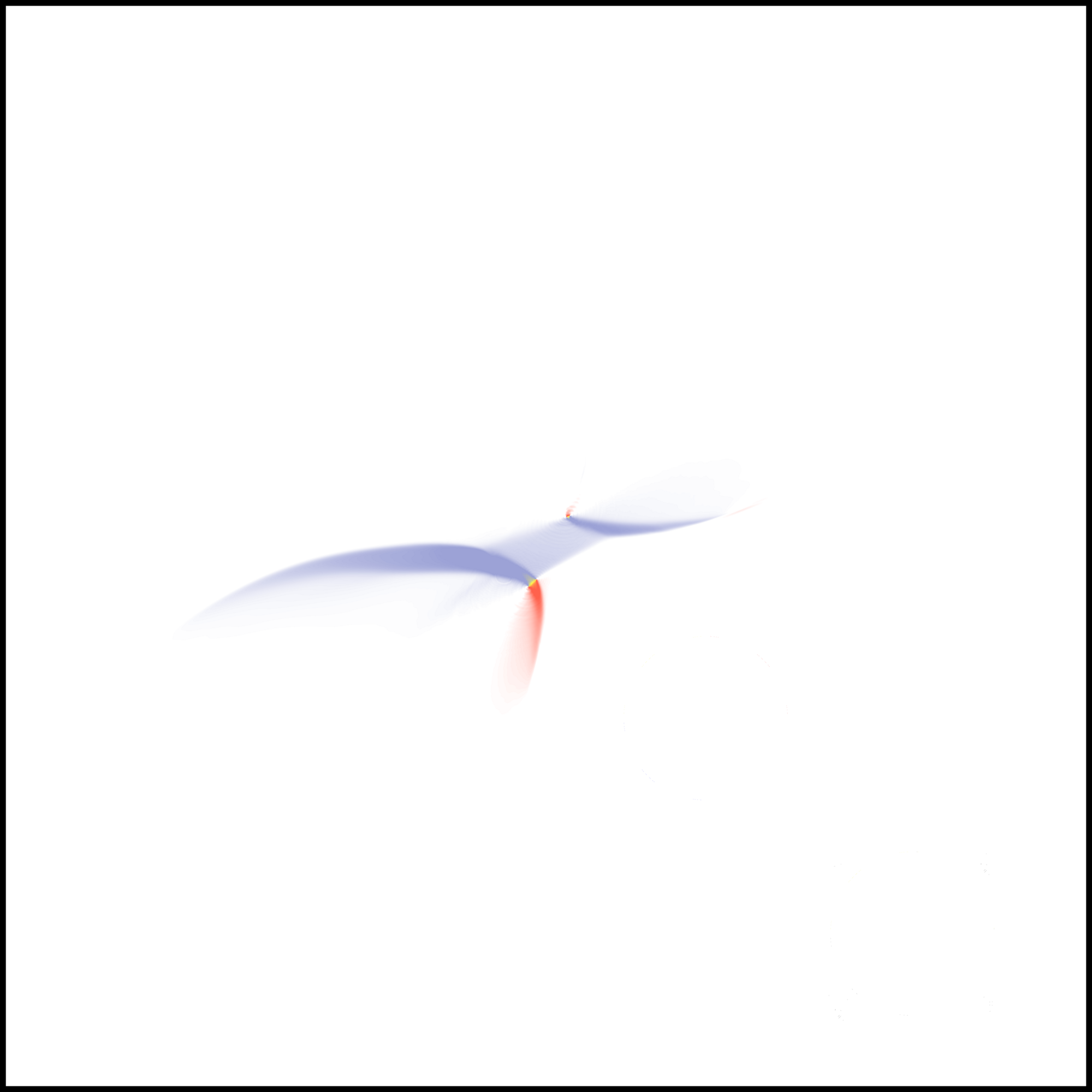} & \includegraphics[width=0.30\linewidth]{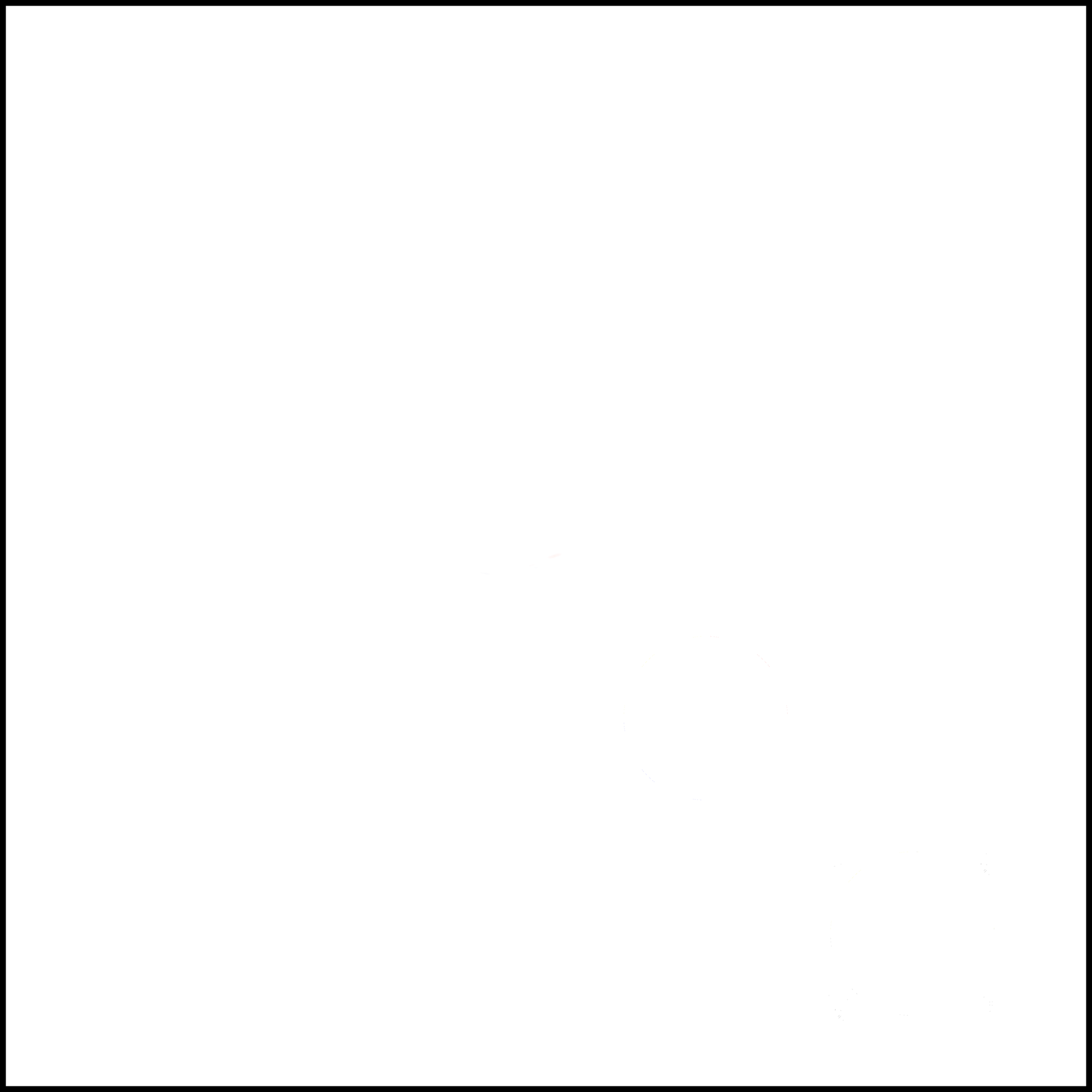}\\
		\end{tabular}
		\caption{The direction of the magnetisation vector in the plane of the simulated sample shown at several times after arrival of the pump pulse.}
		\label{simulations}
		\end{figure}

	The simulated temporal evolution of the switched area of magnetization is shown in Fig.~\ref{simulations}, showing that the switched region relaxes back to its equilibrium state within $\SI{1}{\nano \second}$. The results of the micromagnetic simulations are qualitatively similar to the experimental results shown in the main text, notwithstanding the difference in length and time scales. The simulated sample dimensions were limited by the requirement to include the exchange interaction in the simulation, necessitating in-plane cell-sizes smaller than $\approx$\SI{15}{\nano\meter}. Increasing the spatial dimensions with such cell sizes imposes intolerably high computational overheads. This limitation on the simulation dimensions in turn reduced the time scale of the process.
		
\clearpage

\section{Experimental data}
\subsection{Decay of the switched area}
\begin{figure}[h]
	    \centering
	    \includegraphics[width=0.75\linewidth]{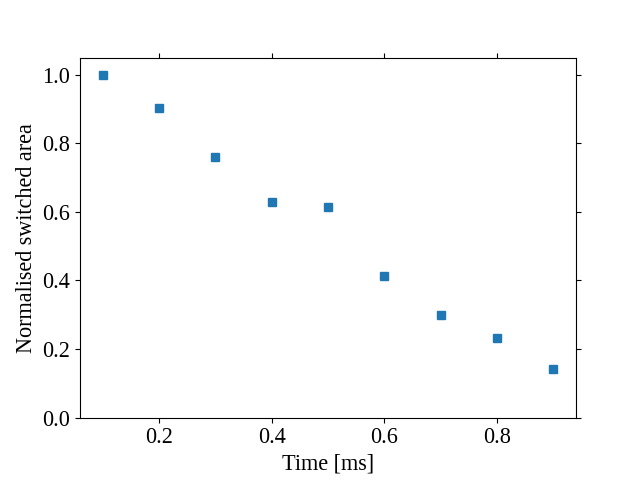}
	    \caption{Graph showing the normalised size of the switched area as function of time, averaged over 20 measurements}
	    \label{decay}
	\end{figure}

\begin{figure}[h]
 	\centering
 	\begin{tabular}{ccc}
 	 	\SI{0.1}{\milli \second} & \SI{0.5}{\milli \second} & \SI{0.9}{\milli \second} \\
 		\includegraphics[width=0.32\linewidth]{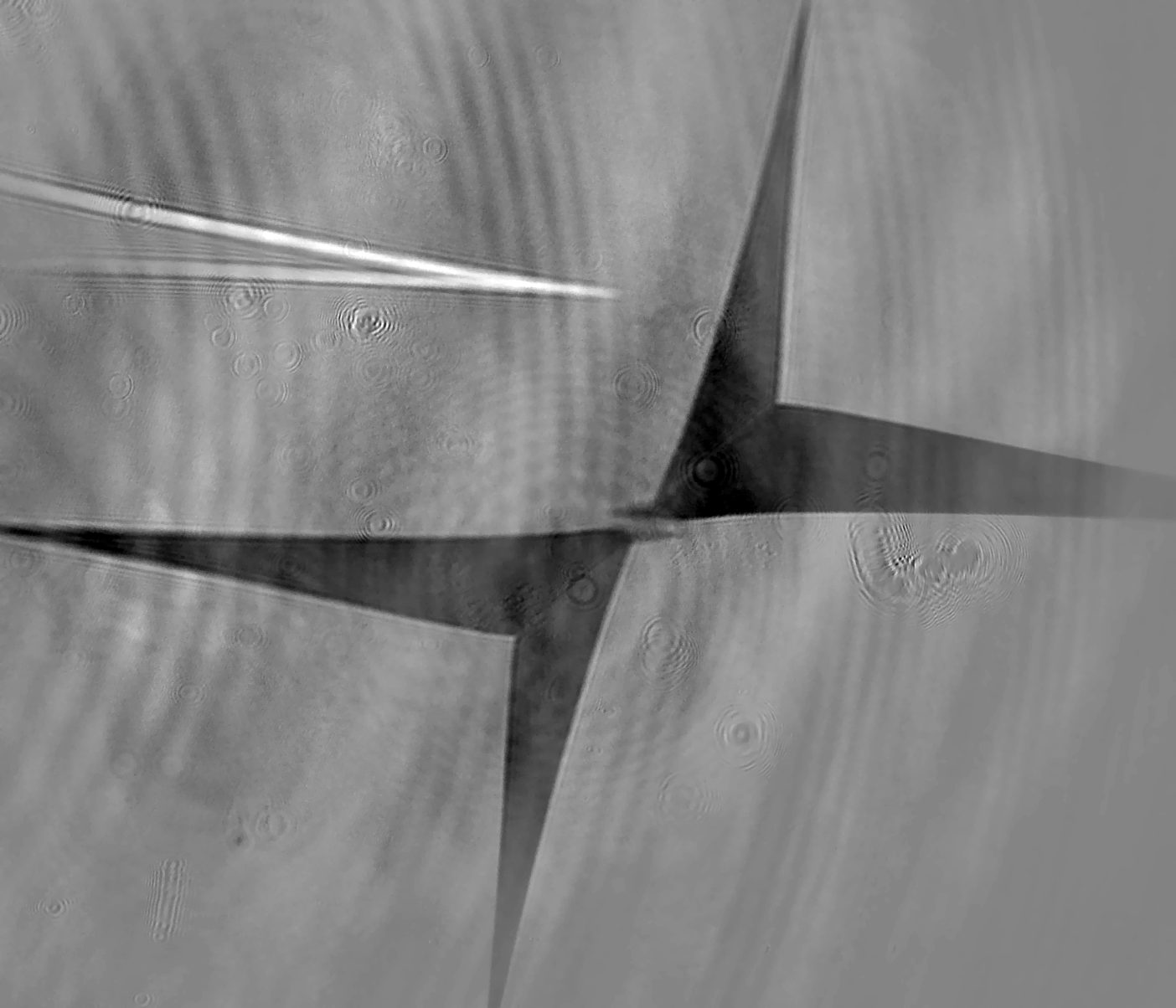} & \includegraphics[width=0.32\linewidth]{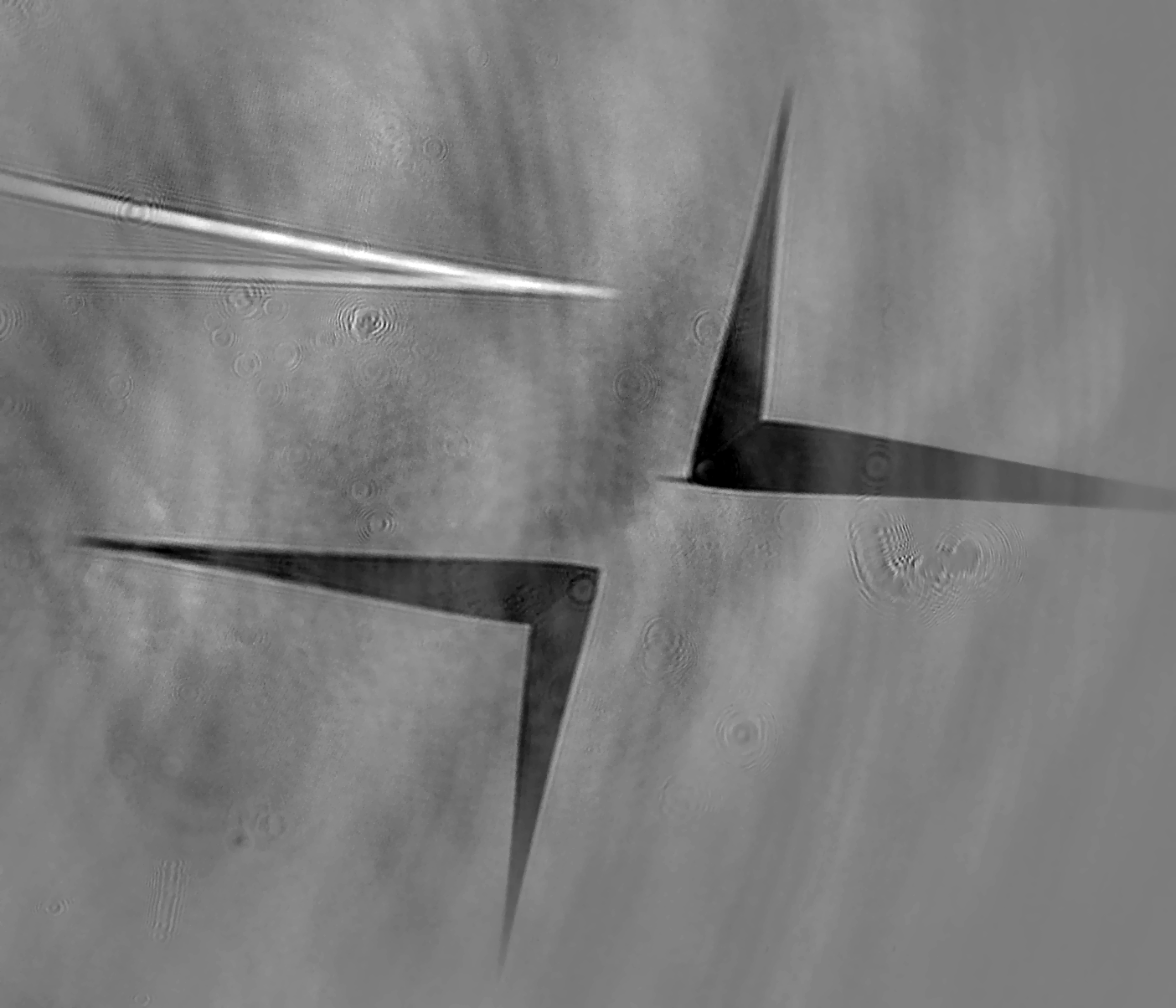} & \includegraphics[width=0.32\linewidth]{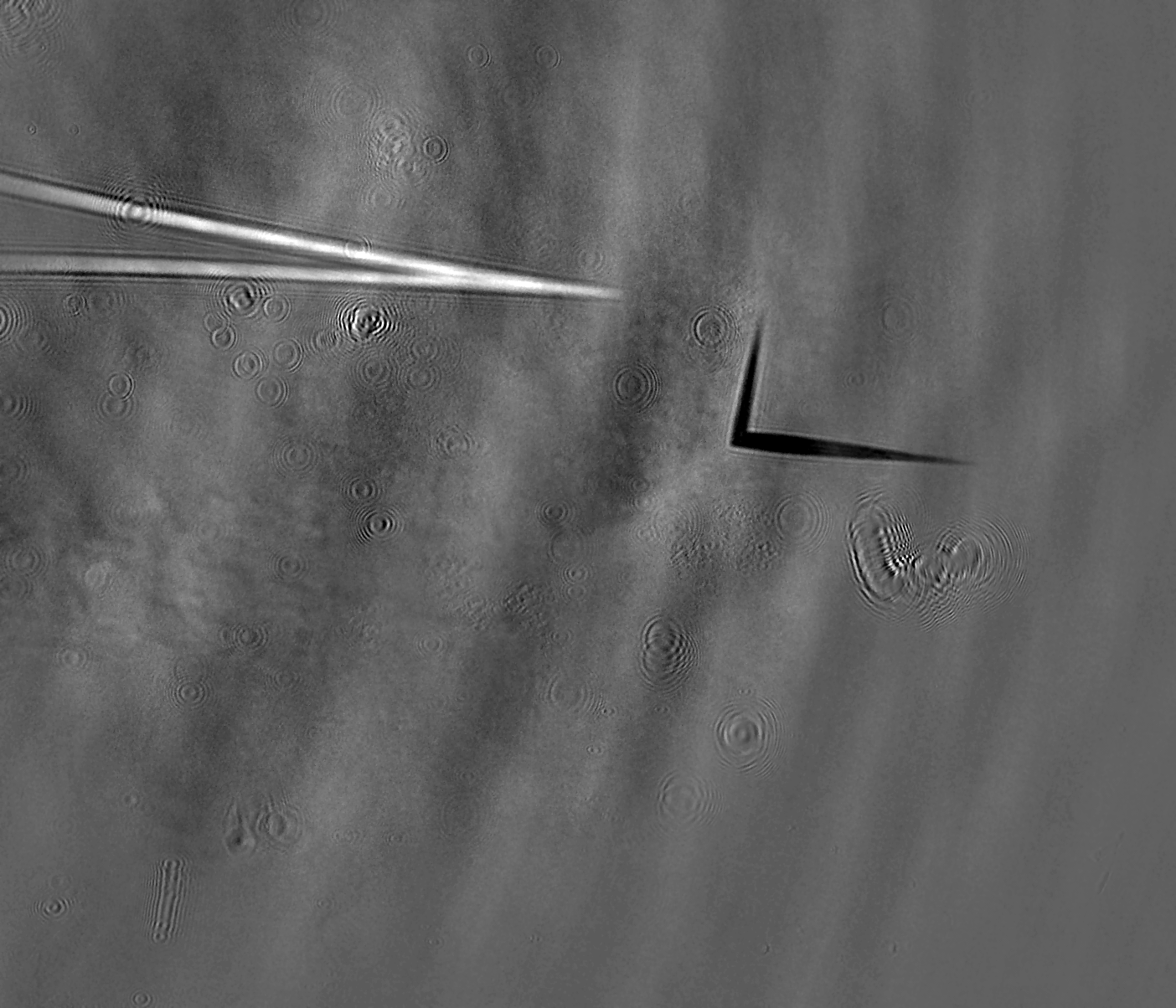}\\
 	\end{tabular}
 	\caption{Millisecond-scale time evolution of the switched pattern, induced by a pump pulse of wavelength \SI{13.0}{\micro \meter} and with an in-plane magnetic field \SI{-0.211}{\milli \tesla}.}
 	\label{decay_pictures}
 \end{figure}

The millisecond-scale decay of the switching was measured, with the results shown in Figs.~\ref{decay} and \ref{decay_pictures}. Decay of the structure happens on the order of a millisecond after being pumped, with some fluctuation between measurements which is ascribed to the fluctuation in pump pulse intensity. Note the complete absence of ripples on this timescale.

\clearpage

\subsection{Spectral dependence of the switching}
\begin{figure}[h]
	    \centering
	    \includegraphics[width=0.7\linewidth]{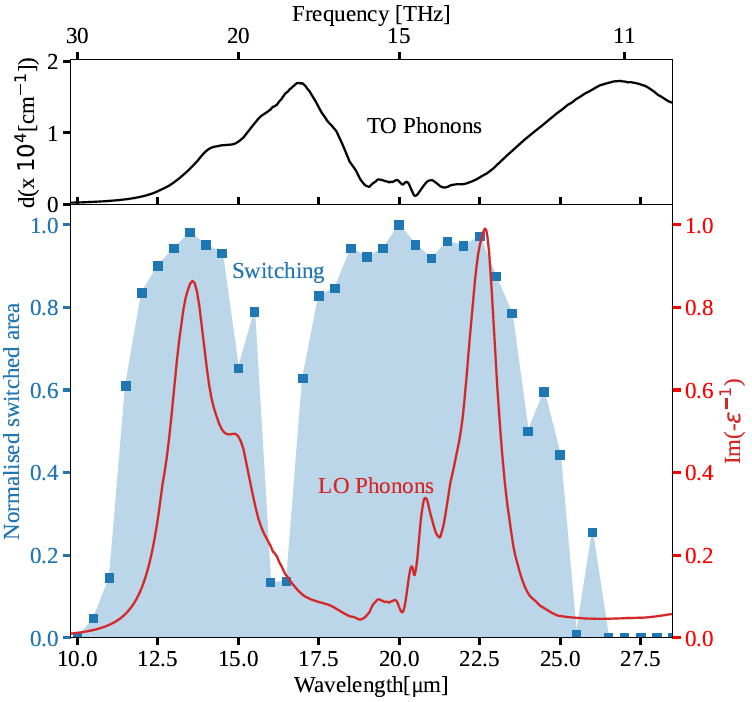}
	    \caption{Graph showing the normalised size of the switched area as function of wavelength averaged over 20 measurements (blue), LO (red) and TO (black) phonon dispersion in YIG. Phonon dispersions taken from Ref.~\cite{Stupakiewicz2021}.}
	    \label{wavelength}
	\end{figure}

The spectral dependence of the switched area, along with the LO and TO phonon dispersions, are shown in Fig.~\ref{wavelength}. This shows the correspondence between magnetic switching and the population of LO phonon modes at a given wavelength. The normalised switched area was calculated by extracting the ratio of the switched area relative to the total image area, followed by normalisation to the wavelength which showed maximal switching. This was then averaged over 20 measurements at each wavelength. The TO and LO phonon dispersions are taken from Ref.~\cite{Stupakiewicz2021}. However, these dispersions were measured for a differently-doped YIG thin film, which could explain why the overlap between switching and LO phonon modes is not perfect.

\clearpage

\subsection{Time dependence for rotated sample}
\begin{figure}[h]
 	\centering
 	\begin{tabular}{ccc}
 	 	\SI{0}{\nano \second} & \SI{50}{\nano \second} & \SI{100}{\nano \second} \\
 		\includegraphics[width=0.27\linewidth]{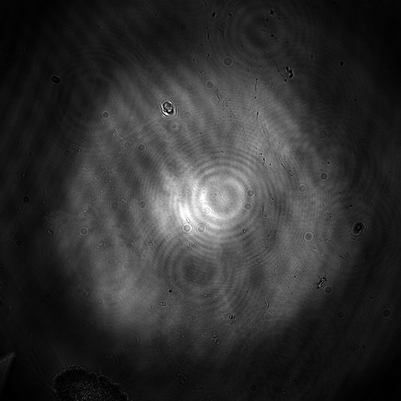} & \includegraphics[width=0.27\linewidth]{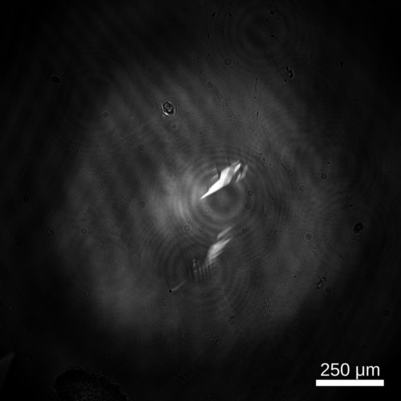} & \includegraphics[width=0.27\linewidth]{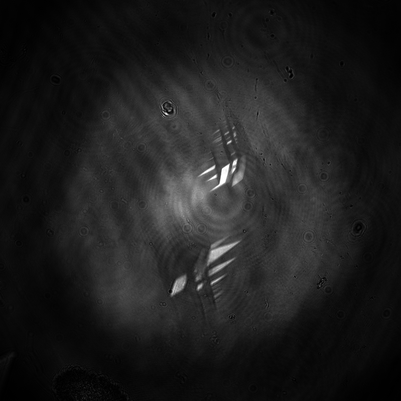}\\
 		\SI{200}{\nano \second} & \SI{300}{\nano \second} & \SI{400}{\nano \second} \\
 		\includegraphics[width=0.27\linewidth]{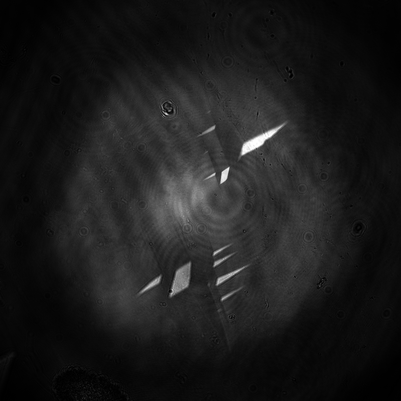} & \includegraphics[width=0.27\linewidth]{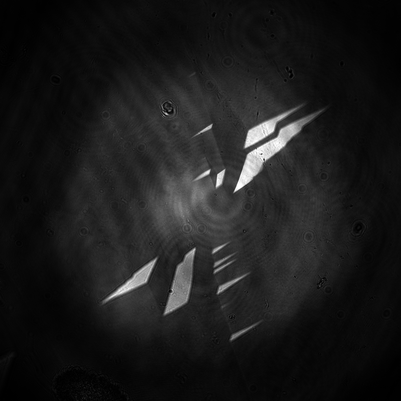} & \includegraphics[width=0.27\linewidth]{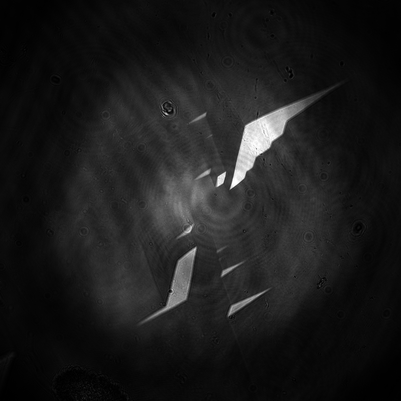}\\
 		\SI{500}{\nano \second} & \SI{750}{\nano \second} & \SI{1000}{\nano \second} \\
 		\includegraphics[width=0.27\linewidth]{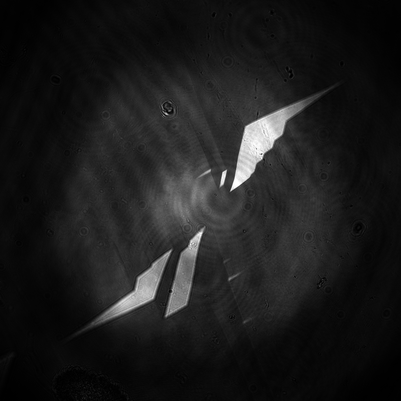} & \includegraphics[width=0.27\linewidth]{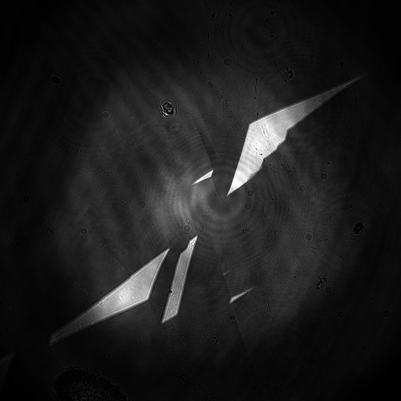} & \includegraphics[width=0.27\linewidth]{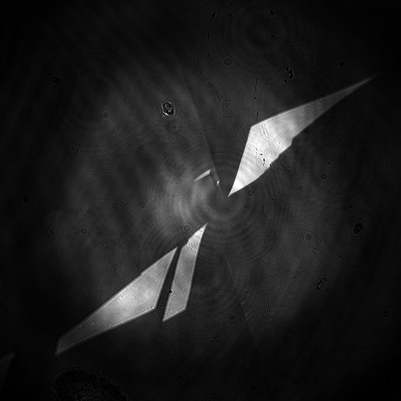}\\
 	\end{tabular}
 	\caption{Time evolution of the switched pattern for a magnetic-field orientation of \ang{-30} relative to that used in the main text. The pump wavelength is \SI{13.0}{\micro \meter} and in-plane magnetic field is of strength \SI{0.033}{\milli \tesla}. Note that the magnetic contrast used during these measurements does not distinguish each magnetisation orientation equally well.}
 	\label{rotated_time_dep}
 \end{figure}

Figure~\ref{rotated_time_dep} shows the time evolution of the switched pattern when the YIG thin film is rotated by \ang{-30} in the plane of the film relative to the rotation used in the main text. By rotating the sample in its plane, the direction of the applied magnetic field is effectively changed. Likewise, the orientation of the polarisation relative to the magnetisation direction of each domain is also altered, changing the observed magnetic contrast. The images shown in Fig.~\ref{rotated_time_dep} were obtained with a pump wavelength \SI{13.0}{\micro \meter} and with an in-plane magnetic field \SI{-0.033}{\milli \tesla}. Note the appearance of ripples during the first hundreds of nanoseconds after the pulse, regardless of the magnetic field's orientation.

 \clearpage
 
 
\end{document}